# #Cyberbullying in the Digital Age: Exploring People's Opinions with Text Mining

**Iman Tahamtan**
University of Tennessee
Knoxville TN, USA
tahamtan@vols.utk.edu

**Li-Min Huang**
University of Tennessee
Knoxville TN, USA
lhuang23@vols.utk.edu

## ABSTRACT
This study used text mining to investigate people's insights about cyberbullying. English-language tweets were collected and analyzed by R software. Our analysis demonstrated three major themes: (a) the major actions that needed to be taken into consideration (e.g. guiding parents and teachers to cyberbullying prevention, funding schools to fight cyberbullying), (b) certain events that were important to people (e.g. the Michigan cyberbullying law), and (c) people's major concerns in this regard (e.g. mental health issues among students). Parents and teachers have an important role in educating, informing, warning, preventing, and protecting against cyberbullying behaviors. The frequency of negative sentiments was almost 2.45 times more than positive sentiments.

## INTRODUCTION
Cyberbullying has been a public concern in the past decade. According to Tokunaga (2010, p. 278), cyberbullying is defined as "*any behavior performed through electronic or digital media by individuals or groups that repeatedly communicates hostile or aggressive messages intended to inflict harm or discomfort on others*". Cyberbullying can negatively impact mental, physical, and behavioral health status of victims and their families (De Bourdeaudhuij, Jacobs, DeSmet, & Gunther, 2015).

The current study aims to reveal people's insights about cyberbullying from tweets. Twitter has become a valuable source of information to explore peoples' behaviors and viewpoints about various public health and social issues (Karami, Dahl, Turner-McGrievy, Kharrazi, & Shaw Jr, 2018). However, few studies (e.g., McHugh, Saperstein, & Gold, 2019) have used Twitter as a data source to explore people's viewpoints on cyberbullying. The present study will contribute to the literature by applying a textual analytics method to reveal people's perception about cyberbullying.

## METHODS
### Data collection
Twitter's Application Programming Interface (API) was applied to collect English-language tweets that contained *#cyberbullying* and *cyberbullying*. Using R software (https://www.r-project.org/) we collected data on April 6th, March 24th, and June 2nd in 2019 and in total retrieved 8,079 tweets, excluding re-tweets.

### Data cleaning and analysis
We cleaned the data (numeric digits, URLs), and removed whitespaces, converted words to lowercase, filtered out stop words and word stemming was done. The screen names that (a) had posted more than one tweet, and (b) tweets with more than 5 hashtags were removed to exclude fake accounts or bots, which resulted in 6,050 tweets. We used Bing's lexicon approach to analyze and understand the negative and positive sentiments. We also performed correlation network analysis to identify which keywords in cyberbullying tweets occurred most often together (see Silge & Robinson, 2017).

## RESULTS
The top 18 words are presented in Table 1. "Cyberbullying" (n=5,932), "bullying" (n=1,066), and "stop" (n=585) were the top three words in the cyberbullying tweets. There were 163 terms in the category of positive sentiment and 391 terms in the category of negative sentiment (Figure 1). The frequency of negative sentiments (n=3,257) was almost 2.45 times more than positive sentiments (n=1,324).

| Word | Counts | Word | Counts |
|---|---|---|---|
| Cyberbullying | 5,932 | Kid | 214 |
| Bullying | 1,066 | Student | 210 |
| Stop | 585 | Media | 209 |
| People | 558 | Social | 206 |
| Law | 419 | Michigan | 198 |
| Don't | 389 | Report | 193 |
| Online | 362 | Video | 187 |
| School | 296 | Internet | 163 |
| Call | 233 | Hate | 60 |

**Table 1. Most common words in cyberbullying tweets**

The correlation network analysis of keywords (Figure 2) revealed seven main clusters of keywords. For example, {experience, children}, {mental health}, {Michigan cyberbullying law}, {girl, photo, share, protect}, and {fund, adequate, school, bullying} were some of these clusters.

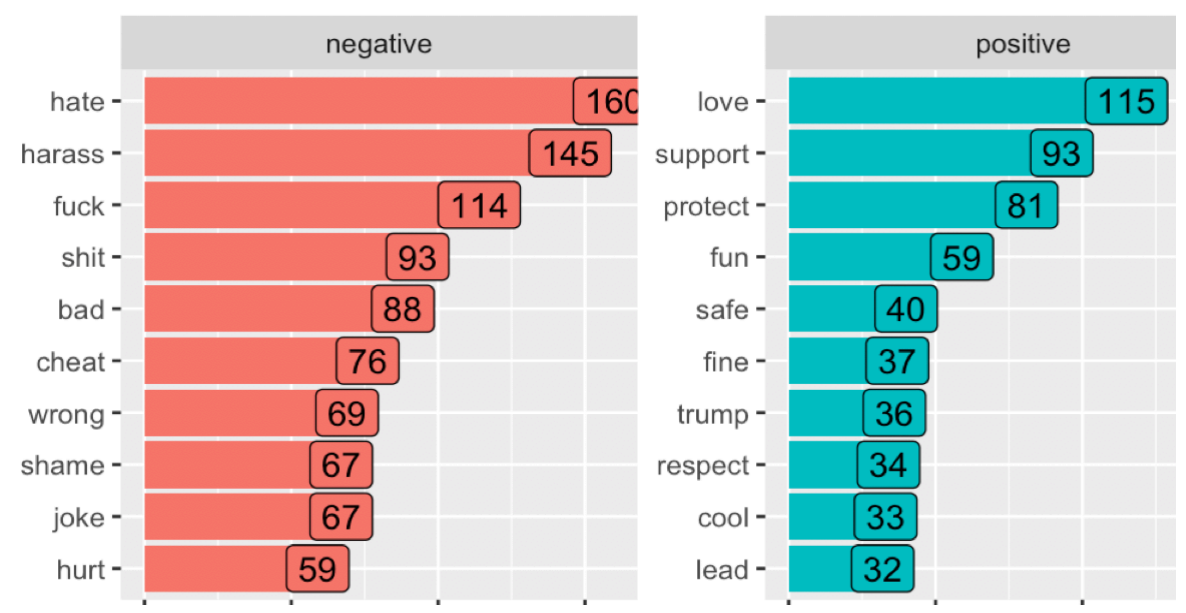

**Figure 1. The frequency of positive and negative sentiments (The number in each chart represents the terms frequency)**

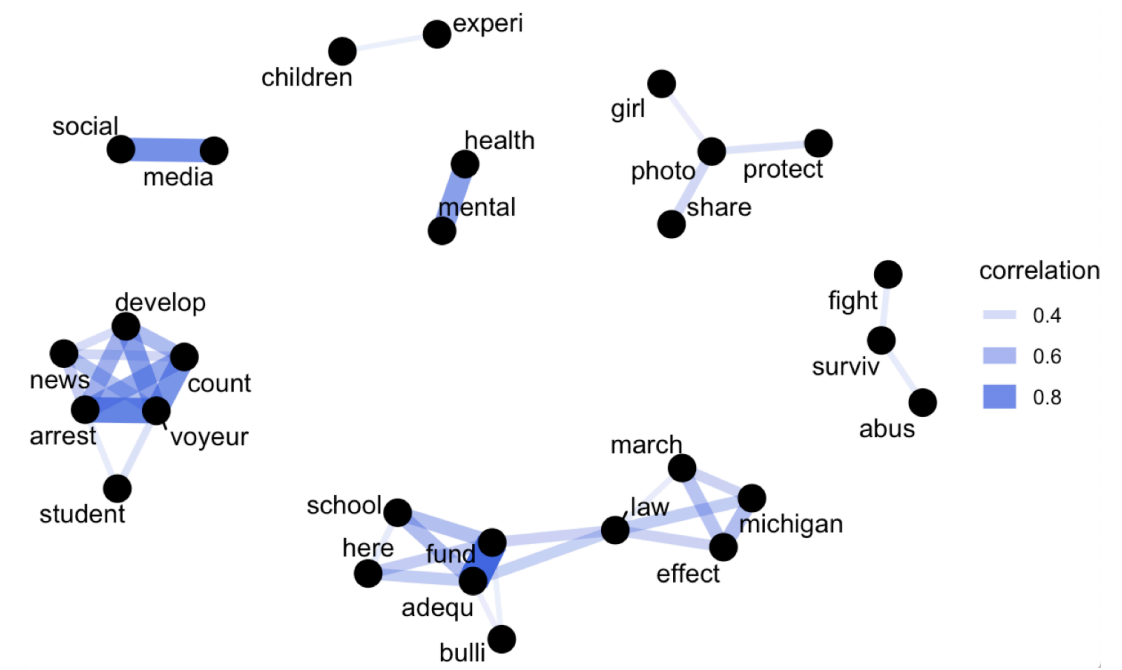

**Figure 2. The correlation network of cyberbullying tweets**

## DISCUSSION

Our analysis demonstrated that "people", "kids", "students", "schools", and "stop" were among the most frequently used words. This might indicate that people are mainly concerned about preventing and stopping cyberbullying behaviors among children and students. The results further revealed that parents and teachers should be trained about the different aspects of cyberbullying, particularly about cyberbullying prevention in schools. One cluster in the correlation network included the terms {fund, adequate, school} which might indicate the importance of dedicating adequate funding to schools to take actions against cyberbullying.

Another cluster included {girl, photo, share and protect}. This could be possibly an indication that gender differences exist in appearance-related cyberbullying. However, this finding needs further investigation. A previous study found that girls are more likely to experience appearance-related cyberbullying, because they tend to post photos or selfies on social media more often than boys. As a result, girls would receive more hurtful comments based on their appearance than boys (Berne, Frisén, & Kling, 2014). In this regard, a study among high school students indicated that female victims were more likely to have attempted suicide because of the depression caused by cyberbullying (Bauman, Toomey, & Walker, 2013).

In addition, many tweets were about the recently launched "Michigan cyberbullying law". This might be representative of the fact that the text mining of tweets could be used to learn certain (or the latest) events that have drawn people's attention. This is in line with Calvin, Bellmore, Xu, and Zhu's (2015) study, who stated that the number of retweets may reflect certain important events. The correlation network analysis demonstrated that people were concerned about the consequences of cyberbullying, mainly the "mental health" issues among kids and students. A few tweets had some advice about how to recognize the signs of cyberbullying. For example, a tweet expressed some of these signs: "*avoiding school or group gatherings, slipping grades, acting out in anger, wanting to stop using your computer or cellphone, and withdrawal from family members, friends and activities*".

## CONCLUSION

Findings suggest that people use social media (Twitter) to share a rich content about cyberbullying-related information and coping strategies from different aspects, including (a) the actions that needed to be taken into consideration, (b) certain events that had drawn people's attention, and (c) people's major concerns in this regard.

## REFERENCES

Bauman, S., Toomey, R. B., & Walker, J. L. (2013). Associations among bullying, cyberbullying, and suicide in high school students. *Journal of adolescence, 36*(2), 341-350.

Berne, S., Frisén, A., & Kling, J. (2014). Appearance-related cyberbullying: A qualitative investigation of characteristics, content, reasons, and effects. *Body image, 11*(4), 527-533.

Calvin, A. J., Bellmore, A., Xu, J.-M., & Zhu, X. (2015). # bully: Uses of hashtags in posts about bullying on Twitter. *Journal of School Violence, 14*(1), 133-153.

De Bourdeaudhuij, I., Jacobs, N. C., DeSmet, A., & Gunther, N. (2015). Comparing associated harm with traditional bullying and cyberbullying: A narrative overview of mental, physical and behavioural negative outcomes *Cyberbullying* (pp. 54-76): Routledge.

Karami, A., Dahl, A. A., Turner-McGrievy, G., Kharrazi, H., & Shaw Jr, G. (2018). Characterizing diabetes, diet, exercise, and obesity comments on Twitter. *International Journal of Information Management, 38*(1), 1-6.

McHugh, M. C., Saperstein, S. L., & Gold, R. S. (2019). OMG U# Cyberbully! An exploration of public discourse about cyberbullying on twitter. *Health Education & Behavior, 46*(1), 97-105.

Silge, J., & Robinson, D. (2017). *Text mining with R: A tidy approach*: " O'Reilly Media, Inc.".

Tokunaga, R. S. (2010). Following you home from school: A critical review and synthesis of research on cyberbullying victimization. *Computers in human behavior, 26*(3), 277-287.